# Modeling and simulation of multiprocessor systems MPSoC by SystemC/TLM2

Abdelhakim ALALI [1], Ismail ASSAYAD[2] and Mohamed SADIK [3]

HASSAN II University, Ecole Nationale Supérieure d'Electricité et de Mécanique,
RTSE team, Route D'El Jadida, Casablanca, Morocco
[1] hakim.alali@gmail.com, [2] i.assayad@ensem.ac.ma, [3] m.sadik@ensem.ac.ma

**Abstract**

The current manufacturing technology allows the integration of a complex multiprocessor system on one piece of silicon (MPSoC for Multiprocessor System-on- Chip). One way to manage the growing complexity of these systems is to increase the level of abstraction and to address the system-level design.

In this paper, we focus on the implementation in SystemC language with TLM (Transaction Level Model) to model an MPSOC platform. Our main contribution is to define a comprehensive, fast and accurate method for designing and evaluating performance for MPSoC systems.

The studied MPSoC is composed of MicroBlaze microprocessors, memory, a timer, a VGA and an interrupt handler with two examples of software. This paper has two novel contributions: the first is to develop this MPSOC at CABA and TLM for ISS (Instruction Set Simulator), Native simulations and timed Programmer's View (PV+T); the second is to show that with PV+T simulations we can achieve timing fidelity with higher speeds than CABA simulations and have almost the same precision.

## 1. Introduction

The literature shows that much of the design time is spent in the performance evaluation. In addition, the iterations in the design flow become prohibitive for complex systems. Therefore, achievement of high performance MPSoCs is a challenge. The solution is strongly linked to the availability of fast and accurate methods for the design and performance evaluation [1] .A modeling approach to reduce the time of design and validation time for MPSoCs is to use the Transaction Level Modeling models (TLM ) [2] at the system level. The SystemC simulation language provides a design and rapid high-level simulation, as opposed to detailed hardware models [11].

In this article, we present a platform MPSOC multi-MicroBlaze, modeled with SystemC 2.2.0 [3]. Some components come from SocLib [4] (open-source library of interoperable models and multi-level SystemC hardware components for modeling and simulation of multiprocessor platforms), and others are components of Xilinx Platform Studio"s library such the MicroBlaze processor, BRAM (Block RAM) with some custom templates.

We adopt a strategy for estimating the performance CABA at several levels, PV (ISS and Native), PV + T. Reference comparison is the CABA level, since it is bit-accurate and cycle around. A main idea widely exposed in the literature is that for better complementarity should be able to choose each time ( even while running ) between either (a) a fast simulation and imprecise , or ( b) a simulation with a increased accuracy but at the cost of longer simulation [5].

Our objectives in this publication are:

• Develop a rapid exploration of performance of design MPSoC tool;

• Show that the PV+T model offers a better alternative than (a) and (b) but at the cost of an additional modeling effort. This latest effort is nevertheless quite acceptable in contrast to the loss of accuracy in (a) or loss of simulation speed in (b). If, despite these losses (a) and are now widely used in system evaluation, this is only because it lacked a better alternative.

The rest of this paper is organized as follows: an overview of related work on existing simulation speedup techniques at TLM for MPSoC is provided in section 2. Section 3 describes the context of use of the platform; Section 4 describes the architecture of the multi-MicroBlaze system. Section 5 presents the simulation platform and the implementation of CABA, ISS, native and PV + T models. Section 6 describes the results of the applications running on the platform.

## 2. Related work

A lot of researches on design exploration and performance evaluation for embedded systems have been conducted. As a result of these researches, several exploration environments are proposed, such as MILAN [6], Metropolis [7], STARSoC [8] and SimSoC [17]. The work presented in this article can be seen as complementary to these environments.

Compared to traditional heterogeneous co-simulation tools, they have not developed an open-source architecture that allows running multiple types of simulation to find the best implementation of the MPSoC in SystemC-TLM in term of performance (speed and accuracy of simulation).

Since the first proposition of TLM in 2000 [9] [10], an increasing number of research projects have considered the problem of its definition, which has led to a multitude of different frameworks [11] [12] [13] [14]. All of these researches have two factors in common: 1) TLM"s are presented as stacks of several levels and 2) the communication and computation aspects of the frameworks are kept separate.

Viaud [15] and al. have proposed an ambitious timed TLM based on conservative parallel discrete event theory. They obtained a high speedup simulation factor but they did not measure this speedup on real applications. Their model is also different from ours. Firstly, our approach can be applied for hierarchical or distributed MPSoC design, and secondly, it is open-source

Kim [16] and Boukhechem [8] propose a new technique for HW/SW co-simulation for heterogeneous MPSoC platforms in timing model PVT, we have all advantages of PVT TA that we refined in order to add it as a priority management. Also we integrated computation and communication simulation.

## 3. Context of use of the platform

A major challenge in designing the architecture of a system is to define the configuration of this architecture. In fact, the designer did not advance a precise idea about the final configuration of this architecture. It is for this reason that it provides a virtual platform for him to explore a set of configurations so that it can make the right choice. For this choice is that it just takes the following two fundamental properties:

- performance evaluated by the virtual platform on the one hand is accurate,
- and secondly, it must be able to choose from a large number of configurations and this is only possible if the virtual platform is fast enough to explore them all in the time available to the designer.

These two properties are intimately related because each impacts the other and therefore should be treated both.

In addition a third fundamental property is also necessary for the designer to make the right choices about the functional correction models configurations. The models of hardware and software components used in these configurations must be consistent in their behavior with the behavior of the physical components of the chip. The correction of the models of software components can be guaranteed only if we ensure that software code running on the virtual platform will run on the chip without modification. Concerning the correction of hardware components it is ensured by a verification approach which is out of the scope of this paper.

## 4. Architecture

The multiprocessor system has a base and a complete architecture. The basic architecture of the platform consists of 2 MicroBlazes each one connected with a 64 KB BRAM via the LMB bus processors. And they are connected to the OPB bus, and a block of 32MB SRAM memory [19]. A high-level view of the architecture of multi-core MicroBlaze is illustrated in Figure 1.

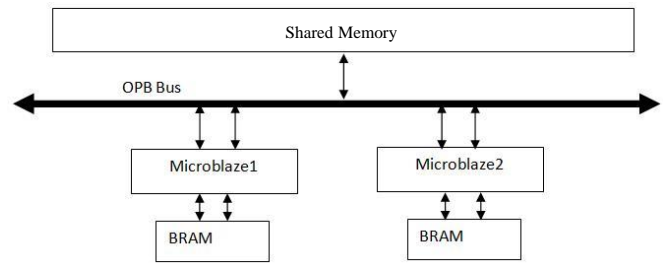

Fig. 1 Basic architecture of multi-MicroBlaze.

In the simulation environment SytemC, the complete design is multi-MicroBlaze implementation of SystemC models, and external test software running on the host PC to stimulate and control the execution of the application on the multiprocessor architecture.

The complete system architecture to simulate consists of 2 MicroBlazes, an interrupt handler, VGA controller, timer, GPIO and SRAM, the model system in SystemC is as follows:

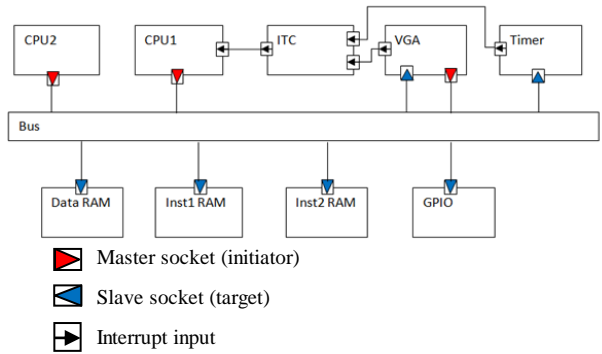

Figure 2. Platform architecture

## 5. SystemC simulation platform

The simulation platform includes multi-MicroBlaze SystemC models for the MicroBlaze processor simulator, BRAM and SRAM. The SystemC components of multi-MicroBlaze system were designed to work together to provide an efficient simulation environment, easy to use and understand. These components are very accurate in time, in accordance with their specifications. Some key features of these models are described below.

Processor Model:

With the TLM approach, the behavior of a processor may have three major descriptions ISS, and PVT Native plus CABA-SystemC which is implemented in the same way that the RTL, most components are SoCLib [7]. In the second description ISS, the processor is modeled with a specific instruction level simulator (ISS: Instruction Set Simulator). Instructions are executed sequentially in this case without reference to the micro-architecture of the component.

We have specified the execution time of each instruction in order to estimate the execution time of the whole application. To implement this description across ISS, we resorted to SoCLib [7], and we have modified the description for PV + T. In the third description, all processors perform application tasks. These tasks are then executed by the machine simulation.

In each sub - level functionality of each processor is disclosed with a module using the SystemC SC_THREAD process. Accordingly, one processor is considered as an active component. Since all components (Timer, VGA, memory ...) are apart from the passive processor, we connected it with the other components as follows: Operations components are executed with the control thread of the processor, when a new transaction is completed, the processor sends a request socket.read () or socket.write () and gets stuck. The thread remains active in the model of bus or memory until it receives the response to the query. Here we are talking about reading or writing way since we use the same path for the request and response. Such an implementation enables the complete simulation acceleration gains because there is no context switch required for the scheduler SystemC. However, with this implementation, we cannot model the components that run simultaneously.

Memory model:

The memory module that we designed is a passive component "slave" type is currency in two parts, one for instructions and one for data and transaction includes two methods: read and write. This structure allows us to accelerate the simulation. These two methods are called and executed directly in the thread initiator connected to the memory component.

In our environment, the target port is connected directly to the bus. The module is shared between two processors. Access time and cycle time parameters are added to the component description to estimate performance.

Bus architecture and model:

Most SoC designs are based on hardware blocks connected together with bus signals, which are classified as groups of data, address, and control links. Several companies provide the following SoC bus architectures so that designers can easily integrate the IP blocks into a single silicon chip: AMBA, Core Connect, CoreFrame, OPB (On-Chip Peripheral Bus), Silicon Backplane Network, and Wishbone. Our architecture platform is designed around the OPB bus. The OPB bus architecture was developed by IBM [21]. It is very simple since it defines only one bus. However, it supports various features depending on the desired bus operations: multiple masters, single cycle read/write, block transfer cycles that systematically perform a set of single read cycles and/or a set of single write cycles. The OPB connect xilinx Microblaze processor.

Moreover, OPB supports various IP block interconnection methods: Up to a 64-bit address bus, 32-bit or 64-bit data bus implementations; Fully synchronous; Provides support for 8-bit, 16-bit, 32-bit, and 64-bit slaves; Provides support for 32-bit and 64-bit masters, Single cycle transfer of data between OPB bus master and OPB slaves; A 16-cycle fixed bus timeout provided by the OPB arbiter.

In this work OPB Bus Master Priority is fixed, priority is set in hardware within the simple arbiter. The system designer assigns relative priorities to OPB master devices via the way they are attached to the arbiter. This is the simplest arbitration procedure. It is the least costly to model and implement.

Table as shown describe the address space for each component.

Table 1: The address space for each

| Component | Start address | size |
|---|---|---|
| *BRAM memory* | 0x00000000 | 0x00002000 |
| *SRAM memory* | *0x20100000* | *0x00100000* |
| *GPIO* | 0x40000000 | 0x00010000 |
| *Interruption controler* | 0x41200000 | 0x00010000 |
| *Timer* | 0x41C00000 | 0x00010000 |
| *VGA controler* | 0x73A00000 | 0x00010000 |

VGA model:

Display controller has a resolution of 640 columns by 480 rows (640 × 480) with a refresh rate of 60 Hertz.

5.1 Types of Simulation

5.1.1 Proposal and justification

To study the performance of MPSOC systems, we need to identify the details of the micro-architecture level to CABA. Especially those related to the communication part and those related to the treatment part.

From a deployment of software architecture MPSOC, we evaluate the performance of our system which allows us to extract the most appropriate solution. At each level of abstraction is a more accurate assessment of performance in less time-consuming simulation. CABA platform model is implemented in SystemC which is the reference memory access and communications are raw signals.

The models below are included in the MPSoC simulator for an exploration of architectures. The methodology for estimating performance must meet the criterion of flexibility to be adaptable to different architectures.

5.1.2 ISS Simulation

Simulators running from the instruction are executable. They decode the bit stream of instructions received by the processor. They are often designed to operate alone and can load the program and manage internal memory [17]. In our case, all external access to processors become transactions in the simulation, these include access all memory access, access to nearby devices if they are simulated as independent components, such as Timer, VGA and interrupt handlers.

The co- simulation of co- application executing on the ISS software in parallel with the rest of SystemC simulation on the hardware platform and a model of the processor CPU in SystemC simulation encapsulates the CPU simulator ISS. All memory access requests for data and instructions are function calls the ISS first. These function calls are transformed into transactions in the TLM model. It is then possible to assess the traffic on the TLM model.

5.1.3 Native simulation

Much of the simulation model calculates the execution processor so that only its outward communications are important. For this purpose, the embedded simulation for native code is compiled for the processor of the computer simulating (HOST), write operations and reading outward are redirected to the simulation [18]. There is no instruction decoding or ISS to run. The simulation is faster. In our project we implemented the native simulation with the total redirection of I / O, which requires the compilation of embedded processors for stimulants software. All memory addresses used in the program are those of the architecture.

5.1.3 PV + T Simulation

Proposed to implement PV + T methodology should also consider issues related to the time synchronization of processors, the dynamic contention in the bus and the specification of the communication protocol [8]. Refining requires a thorough study of each function to derive a precise execution time. However, it is not necessary to refine all the functions of MPSoC to simulate.
The idea is to describe the temporal and the new granularity of communications in separate Timing model information that can be seen as a particular aspect of a kind [10]. The estimated performance level PV + T returns to evaluate performance of two parts calculation and communication time.
In our case, to assess the time of each task we used the simulator MicroBlaze processor ISS level but adding time.

For this we mainly identified the number and type of instructions executed as relevant activities in the processor component.

Turnaround instructions from MicroBlaze processor are estimated from the technical documentation provided by. Below is an example of our thread implementation to implement the functionality of the calculation part (processor) described in sub-level PV + T:

```
void MicroBlazeIss::step(void) {
IDecode(m_ir, &ins_opcode, &ins_rd, &ins_ra, &ins_rb, &ins_imm);
switch (ins_opcode) {

case OP_ADD:
next_pc = r_npc + 4;
Wait(ADD_delay,sc_core::SC_NS)
          break;
……….
 case OP_LW:
…..
        LOAD(READ_WORD,
        addr); next_pc = r_npc + 4;
        Wait(Transaction_delay,sc_core::SC_NS)
        break;
………….
case OP_SB:
……….
        STORE(WRITE_BYTE, addr,data);
        next_pc = r_npc + 4;
        Wait(Transaction_delay,sc_core::SC_NS);
                                break;
  }
}
```

Fig. 3. Calculation part of processor

It is noted that the instruction execution time MicroBlaze processor are estimated from the technical documentation given by [19].

In this paper we identify the steps needed to run the software instruction level. The processor begins with a reading phase of the next instruction from the instruction memory initializing a request m_iss.getInstructionRequest ( ins_asked , ins_addr ) . The ins_addr parameter specifies the address of the instruction, ins_asked represents the state of the application (or not made) variable. The second step is to decode the instruction in microblaze.cpp to identify the type of the operation via the IDecode function(m_ir , & ins_opcode , & ins_rd , & ins_ra , & ins_rb , & ins_imm) . The next step is reading the operands from memory by tlm:: tlm_response_status stat = socket.read (ins_addr , localbuf) . The final phase involves the execution of the current instruction and updates the processor registers and the program counter.

## 5.2 Software integration

We have two applications were tested in the platform: a game of life and adder integers:

- The game of life: The universe of the Game of Life is an infinite two-dimensional orthogonal grid of square cells, each of which is in one of two possible states, alive or dead. Every cell interacts with its eight neighbors, which are the cells that are horizontally, vertically, or diagonally adjacent [14].

- Adder: The choice of this function simply tests the functionality of the processor.

## 6. Results and discussion

Several experiments were conducted using the same applications and configurations MPSoC system to evaluate CABA levels, ISS, Native and PV + T. In the model of time (PV + T), we integrate specifications OPB and the time between events as the specifications for the MicroBlaze protocol.

To calculate the speedup of the simulation we implemented a function to calculate the start time and the end of the simulation: t2 = "end time", t1 = "start time".
Speed-up formula:

$$\frac{(t2 - t1)_x}{(t2 - t1)_{bit}}$$

Precision formula:
Δx -Δbit
With x = PVT,ISS or Native.

Table 2: simulation results for CABA, ISS, Native and PV+T

| Input | Simulation Type | Speed-up | precision |
|---|---|---|---|
| **adder** | PV+T | 4 | 2% |
| | ISS | 3 | 3 |
| | Native simulation | 15 | 15% |
| | CABA | 1 | 0% |
| **Game of life** | PV+T | 11 | 3% |
| | ISS | 9 | 6% |
| | Native simulation | 102 | 25% |
| | CABA | 1 | 0% |

The experimental results show that the adoption of SystemC as development language and TLM as modeling approach at high levels of abstractions design, can significantly reduce the time of design validation, and allow the development of models very quickly. In addition, the simulation results at higher levels of abstraction show:

- For the ISS and ISS + T, there are no communication costs between ISS C model and its wrapper SystemC, and precisely the ISS + T approach has minimum error accuracy for all tested configurations while having a good acceleration factor.

- Native model is very fast in terms of simulation speed but has a precision error indicating that this model would be very useful for HW / SW functional co-simulation of large SoC based on RISC processors.

The precision error with PV+T is minimal for all configurations tested and has a good acceleration factor. We believe that the use of a new model (PV + time + priority) that integrates event-based priorities management between the two processors transactions, may be obtained by adding these priorities to the PV+T level of simulation and is likely to minimize the errors in the estimations of PV+T. Compared to the sub-level ISS, the level PV + T slows the simulation by 30%.

A precise analysis of the trace produced by the SystemC simulator shows that 80% of the simulation time is made for the execution of the function of the bus while the simulation time of the calculation part is almost zero which reflects our choice to treat the case of PV + T + P.

Also, we noticed that the nature of the software running on the platform impacts performance differences between the four levels. Thus using our platform, we could choose an acceptable level and significantly reduce the development effort compared to CABA level.

## 7 Conclusions

In this paper, we describe the systems at the transaction level for ISS, Native and PV + T, in this latter case is our implementation approach in the sense to estimate the performance in terms of acceleration (simulation time) and the precision of the simulation of systems MPSoC. Different material components have been designed to implement the three levels. To obtain an accurate prediction of the execution time in our environment, we have enriched the level PV + T by timing patterns to an estimation error on the accuracy of the system description.

As future work, we plan to develop PV+T+P model in our platform.